\documentclass[structabstract]{aa}

\usepackage{graphicx}
\usepackage{txfonts}

\newcommand{\etal}{{\em et al.\ }}

\def\ion[#1 #2]{#1\,{\sc #2}}
\def\ergs[#1]{#1 {ergs}~{cm$^{-2}$}\,{s$^{-1}$}\,{sr$^{-1}$}}
\def\dens[#1]{10$^{#1}$\hskip 1.5pt{cm$^{-3}$}}
\def\densr[#1 #2]{10$^{#1}$\hskip 1pt{--}\hskip .5pt{10$^{#2}$}\hskip 1.5pt{cm$^{-3}$}}
\def\fl[#1 #2]{{#1}$\pm${#2}}
\def\orb[#1 #2]{{$#1^{#2}$}}
\def\ls[#1 #2]{{$^{#1}${#2}}}
\def\tm[#1 #2 #3]{{$^{#1}${#2}$_{#3}$}}

\begin{document}

\title{Monte Carlo Markov Chain DEM reconstruction of isothermal plasmas}

\author{E. Landi\inst{1} \and F. Reale\inst{2}\inst{3} \and P. Testa\inst{4}}

\institute{Department of Atmospheric, Oceanic and Space Sciences, University of Michigan, Ann Arbor, MI 48109\\
\and Dipartimento di Scienze Fisiche ed Astronomiche, Sezione di Astronomia, Universita' di Palermo, Piazza del Parlamento 1, 90134, Italy\\
\and INAF-Osservatorio Astronomico di Palermo, Piazza del Parlamento 1, 90134 Palermo, Italy\\
\and Smithsonian Astrophysical Observatory, 60 Garden St., Cambridge MA 02138, USA}

\abstract
{Recent studies carried out with SOHO and Hinode high-resolution spectrometers have shown 
that the plasma in the off-disk solar corona is close to isothermal. If confirmed, these 
findings may have significant consequences for theoretical models of coronal heating. 
However, these studies have been carried out with diagnostic techniques whose ability to
reconstruct the plasma distribution with temperature has not been thoroughly tested.}
{In this paper, we carry out tests on the Monte Carlo Markov Chain (MCMC) technique
with the aim of determining: 1) its ability to retrieve isothermal plasmas from a set of spectral
line intensities, with and without random noise; 2) to what extent can it discriminate between 
an isothermal solution and a narrow multithermal distribution; and 3) how well it can detect 
multiple isothermal components along the line of sight. We also test 
the effects of 4) atomic data uncertainties on the results, and 5) the number of ions whose
lines are available for the DEM reconstruction.}
{We first use the CHIANTI database to calculate synthetic spectra from different 
thermal distributions: single isothermal plasmas, multithermal plasmas made of multiple 
isothermal components, and multithermal plasmas with a Gaussian DEM distribution with 
variable width. We 
then apply the MCMC technique on each of these synthetic spectra, so that the ability 
of the MCMC technique at reconstructing the original thermal distribution can be 
evaluated. Next, we add a random noise to the synthetic spectra, and repeat the 
exercise, in order to determine the effects of random errors on the results. We 
also we repeat the exercise using a different set of atomic data from those used
to calculate synthetic line intensities, to understand the robustness of the results 
against atomic physics uncertainties. The size of the temperature bin of the MCMC 
reconstruction is varied in all cases, in order to determine the optimal width.}
{We find that the MCMC technique is unable to retrieve isothermal plasmas to better
than $\Delta \log T \simeq 0.05$. Also, the DEM curves obtained using lines 
calculated with an isothermal plasma and with a Gaussian distribution with FWHM 
of $\log T \simeq 0.05$ are very similar. Two near-isothermal components can be 
resolved if their 
temperature separation is $\Delta \log T=0.2$ or larger. Thus, DEM diagnostics 
has an intrinsic resolving power of $\log T=0.05$. Atomic data uncertainties may
significantly affect both temperature and peak DEM values, but do not alter our
conclusions. The availability of small sets of lines also does not worsen the
performance of the MCMC technique, provided these lines are formed in a wide
temperature range.}
{Our analysis shows the present limitations in
our ability to identify the presence of strictly isothermal plasmas in stellar 
and solar coronal spectra.}


\maketitle

\section{Introduction}

Detailed measurements of the temperature distribution of the plasma of solar and stellar
coronae are  of critical importance to understand how they are heated to 
multimillion degree temperatures, so it is no surprise that the thermal structure of the solar
corona has been studied ever since the publication of the seminal paper of Edlen (1942), who
discovered its high temperature for the first time. These studies have known a revival in 
recent years, thanks to the availability of a large body of observations obtained with the 
new instrumentation carried on board the SOHO, TRACE, STEREO, Hinode and SDO satellites. 

Plasma loops are a fundamental component of the solar corona, since they are ubiquitous
both in quiet and active solar plasmas. Steady-state models of loops predict that the 
temperature of these structures varies along the loop (Rosner \etal 1978), and that
loops with different length can have very different temperatures; nanoflare-based models 
predict finely-structured loops filled with multithermal plasma all along their length 
(Patsourakos \& Klimchuk 2005). A consequence of these models is that quiet coronal 
plasmas observed {\em outside the solar limb} should be multithermal because they are the 
sum of a multitude of loops with different lengths and temperatures along the line of 
sight. The question of clearly distinguishing between multithermal and isothermal plasma
along the line of sight is thus very important in the framework of coronal heating models
(e.g. Klimchuk 2006, Reale 2010). However,
recent measurements have provided increasing evidence of a nearly isothermal nature of 
both coronal hole and quiet Sun plasmas when they are observed above the solar limb,
contrary to model predictions. Several authors (e.g. Feldman \etal 1998 and 1999, Warren 
1999, Doschek \etal 2001, Landi 2008) reported single-temperature determination which 
are remarkably similar even if taken with different instruments and at different times 
during the solar cycle. A constant temperature was also found to characterize an entire 
streamer by Landi \etal (2006). Even a complex active region, when observed at the solar 
west limb, seemed to be composed by three near isothermal plasmas
(Landi \& Feldman 2008). Further studies carried out with Hinode/EIS observations, on the
contrary, provided evidence that the quiescent solar corona is not {\em strictly} 
isothermal, but it is characterized by a rather narrow temperature structure with a tail 
that extends to higher temperatures (Warren \& Brooks 2009, Brooks \etal 2009, Hahn \etal 2011).
The aim of the present paper is to assess how reliable such isothermal plasma claims are.

These measurements have been carried out either with the Emission Measure (EM) loci 
technique or with traditional Differential Emission Measured (DEM) techniques (see
Feldman \& Landi 2008 and references therein). The former allows to determine the temperature 
and the total emission measure (EM) of an isothermal plasma, and it has been the technique 
used in most of the studies that resulted in the isothermal corona results. This technique,
while useful in the case of strictly isothermal plasmas, fails at providing an assessment
of the uncertainty of its results. Traditional DEM techniques rely on several different 
methods to reconstruct the thermal structure of multithermal plasmas determining the 
DEM curve. However, most of them rely on 
assumptions on the smoothness of the distribution, that make it very difficult to 
study isothermal or near-isothermal plasmas. 

Landi \& Klimchuk (2010) carried out tests aimed at evaluating the ability of the EM 
loci technique to discriminate between isothermal and multithermal plasmas, when the 
spectral line intensities used for the reconstruction included uncertainties. They 
also developed a quantitative method to determine how multithermal could a plasma 
distribution be in order to be compatible with a set of observed line intensities 
that was also compatible with an isothermal condition.

In the present work we extend the Landi \& Klimchuk (2010) study to the application
of DEM diagnostic techniques to spectral line intensities. In particular, we aim 
at: 1) determining the ability of such techniques to retrieve isothermal plasma from
a set of spectral line intensities, with and without random noise; 2) understanding 
to what extent are DEM diagnostic techniques able to discriminate between a truly 
isothermal plasma and a multithermal plasma with a narrow distribution; 3) quantifying
how well can such technique detect and resolve multiple isothermal components along
the lines of sight; 4) investigate the effects of atomic data uncertainties on
the results; and 5) study how DEM reconstructions are affected by the number of ions
whose lines are available.

Similar studies with different goals were recently carried out in preparation for the 
SDO and Hinode missions by Golub \etal (2004) and Weber \etal (2004). Both studies 
were focused on determining how well,
and with what uncertainties, could a DEM curve known {\em a-priori} be reconstructed
using data from the limited number of broad-band filter intensities provided by
Hinode/XRT and narrow-band filters in SDO/AIA. They also discussed how efficiently 
could DEM curves from the large number of pixels expected from such images be 
determined and visualized. Unlike the present work, such studies were aimed at 
imaging instruments and did not use
spectral lines intensities; also, no guidelines were provided with regard to the 
interpretation of almost isothermal results. Further, the results of such tests strongly
depend on the specific temperature response function of each instrument, so that they can
not be easily extended to other instruments. 

Line intensities can potentially provide 
better constraints on DEM curves due 1) to the better temperature resolution provided 
by their emissivities; and 2) to the finer sampling of the temperature range of the 
solar corona provided by the large number of ions included in the spectral range
of the available spectrometers (CDS and SUMER on SOHO, EIS on Hinode). In the
present work, we focus on spectral lines; given the larger amount of available
constraints, we can hopefully obtain further insight on the intrinsic abilities
of DEM diagnostic techniques at studying isothermal or near-isothermal plasmas.
Our results can be applied to analyses of spectra observed by any of the available 
spectrometers observing the solar upper atmosphere, such as Hinode/EIS, SOHO/CDS,
SOHO/SUMER, SOHO/UVCS, EUNIS, SERTS, HRTS and, to a certain extent, even the lower-resolution
SDO/EVE. In fact, 
different spectrometers working at different wavelength ranges may still observe 
lines from the same element, so that they will be sampling the same temperature 
interval regardless of their passband. Thus, our results will be of more 
general use than Golub \etal (2004) and Weber \etal (2004).

Many different DEM diagnostic techniques have been developed in the past (see the
reviews of Phillips \etal 2008 and Harrison \& Thompson 1992). Harrison \& Thompson
(1992) carried out a comparative analysis aimed at determining which, among six
different DEM diagnostic techniques, was most successful at reproducing pre-defined
DEM curves using a set of emission lines. More recently, Mark Weber and Paul Boerner
led a similar study in preparation to the SDO mission ({\em http://www.lmsal.com/~boerner/demtest/}),
where, however, the results are not clearly discussed and no clear conclusions are
drawn; also, such a study does not appear to have been published in the literature.
Most importantly, Boerner \& Weber did not address the ability at discriminating 
between isothermal and non-isothermal plasmas, which is the focus of the present 
work. 

After inspecting the material in the Boerner \& Weber website, we found that
Monte Carlo Markov Chain (MCMC) method developed by Kashyap \& Drake (1998) provided
the most robust and accurate results. Also, this method dispenses from many of the
assumptions common to other DEM diagnostic techniques, and provides an assessment
of the uncertainties of the final DEM curve. Thus, we chose to focus on this method in 
the present study. Kashyap \& Drake (1998) and Kashyap \etal (2004) briefly described 
tests made to ensure and assess the robustness and reliability of the MCMC technique, 
but they did not focus
on the ability of the MCMC technique to discriminate between an isothermal
and multithermal plasma. Here we also provide a set of empirical guidelines for the 
choice of the input parameters and for the interpretation of MCMC reconstructions.

Section~\ref{method} introduces the main plasma structure diagnostic techniques, including
MCMC, and discusses their assumptions; it also explains the methodology we follow in our 
tests. Section~\ref{results} describes the results of our tests for several different ad-hoc 
distributions, and Section~\ref{conclusions} reviews the results of this study.

\section{Analysis method}
\label{method}

\subsection{Emission line intensities}

Solar coronal plasmas are usually tenuous enough to be optically thin below 2000~\AA.
The intensity of an optically thin emission line can be written as

\begin{eqnarray}
I = \frac{1}{4\pi d^2}\int_{V}^{}G{\left({T,N_e}\right)}N_{e}^{2}dV
\label{intensityintegral2}
\end{eqnarray}

\noindent 
where $N_{e}$ is the electron density, $V$ is the emitting volume along the line of 
sight, $d$ is the distance between the emitting source and the observer, and 
$G{\left({T,N_e}\right)}$ is the {\em Contribution Function} of the emitting line 
defined as

\begin{equation}
{G{\left({T,N_e}\right)}} = {N_j{\left({X^{+m}}\right)}\over{{N{\left({X^{+m}}\right)}}}}~{N{\left({X^{+m}}\right)}\over{{N{\left({X}\right)}}}}~{N{\left({X}\right)}\over{{N{\left({H}\right)}}}}~{N{\left({H}\right)}\over N_e}~{A\over N_e}
\label{giditi}
\end{equation}

\noindent
where

\begin{itemize}
\item ${N_j{\left({X^{+m}}\right)}\over{N{\left({X^{+m}}\right)}}}$ is the relative 
population of the upper level $j$ and depends on electron temperature and density;
\item ${{N{\left({X^{+m}}\right)}}\over{N{\left({X}\right)}}}$ is the relative abundance
of the ion $X^{+m}$ ({\em ion fraction}); under ionization equilibrium conditions, it 
depends on the electron temperature;
\item ${{N{\left({X}\right)}}\over{N{\left({H}\right)}}}$ is the abundance of the element
$X$ relative to hydrogen;
\item ${N{\left({H}\right)}\over N_e}$ is the hydrogen abundance relative to the electron
density ($\approx 0.83$ for fully ionized plasmas);
\item ${A}$ is the Einstein coefficient for spontaneous emission.
\end{itemize}

\noindent
When the plasma is multithermal Equation~\ref{intensityintegral2} can be
rewritten by defining the {\em Differential Emission Measure} (DEM),
$\varphi{\left({T}\right)}$, as

\begin{equation}
I = \frac{1}{4\pi d^2}\int_{V}^{}G{\left({T,N_e}\right)}\varphi{\left({T}\right)}dT 
\quad \mathrm{with} \quad
\varphi{\left({T}\right)} = N_e^2\frac{dV}{dT}
\label{dem1}
\end{equation}

\noindent
The DEM indicates the amount of material in the plasma as a function of
temperature. When the plasma is isothermal at temperature $T_c$, we can 
define the {\em Emission Measure} (EM) of the plasma as

\begin{equation}
I = \frac{1}{4\pi d^2}G{\left({T_c,N_e}\right)}EM 
\quad \mathrm{with} \quad
EM = \int_V{N_e^2dV}
\label{em2}
\end{equation}
\noindent
The EM of the plasma is a measure of the total amount of plasma
at the temperature $T_c$.

\subsection{Thermal structure diagnostic techniques}

When the plasma is isothermal at the temperature $T_c$, the EM can be 
determined from line intensities as:

\begin{eqnarray}
EM = 4\pi d^2 \frac{I}{G(T_c,N_{e})}
\end{eqnarray}

\noindent
The {\em EM loci} diagnostic technique allows to measure simultaneously the
plasma $EM$ and $T_c$ values. It consists of calculating for each line the 
function $EM(T)$ defined as

\begin{eqnarray}
EM(T) = 4\pi d^2 \frac{I}{G(T,N_{e})} \quad \longrightarrow \quad EM(T_c) =  EM
\end{eqnarray}

\noindent as a function of electron temperature, using the observed
line intensities $I$. Since the plasma $EM$ is the same
for all lines, all the $EM(T)$ curves, when displayed as a function of
$T$ in the same plot, should cross the same point coordinates $(T_c,EM)$. 
The presence of this crossing point also confirms that the plasma is 
isothermal. When the plasma is not isothermal, the $EM(T)$ curves do 
not cross the same point and the EM loci technique can not be used. 

When the plasma is multithermal, the plasma DEM needs to be determined. 
There are many different types of DEM diagnostic techniques, and they 
have been reviewed by Phillips \etal (2008). Some techniques rely on 
the inversion of Equation~\ref{dem1}, others on an iterative procedure; 
Monte Carlo methods are also available. 

The inversion of Equation~\ref{dem1} is carried out by discretising the 
temperature interval in bins whose width is chosen by the user. In 
each bin the DEM is assumed to be constant so that the observed intensity 
$I_s$ of the spectral line $s$ can be written as

\begin{equation}
I_s = \frac{1}{4\pi d^2} \sum_{i=1}^{N} \varphi_i \int_{T_i}^{T_{i+1}}{G_s{\left({T}\right)}dT} 
\end{equation}

\noindent
The values of $\varphi_i$ are determined minimizing via an iterative technique 
the sum the quantity $H=S+\chi^2$, where S is the {\em entropy} of the the 
quantity $\varphi_i$ and $\chi^2$ has the standard definition. In this procedure, 
each observed line is arbitrarily associated to a temperature value $T_{eff}$ 
(discussed below), and the results for all lines 
whose $T_{eff}$ falls in the same bin are combined into a single value. The DEM 
curve is then determined by fitting the $\varphi_i$ values of all temperature bins 
with a polynomial or a spline function.

Iterative techniques start from an arbitrary initial DEM curve and use each 
spectral line to calculate a correction $C{\left({T_{eff}}\right)}$ associated 
to a temperature $T_{eff}$ (discussed below). $C{\left({T_{eff}}\right)}$ 
values whose $T_{eff}$ lies in the same temperature bin (whose arbitrary width 
is chosen by the user) are averaged together. The resulting average corrections 
for all bins are then interpolated with a polynomial or a spline function and 
applied to the initial DEM curve to calculate the new curve to be used in the 
next iteration as initial DEM.

The definition of $T_{eff}$ for each line is rather arbitrary. In many cases 
$T_{eff}$ is assumed to be $T_{max}$, the temperature of maximum abundance of
the ion emitting the line. In other cases, it is defined as some sort of 
DEM- and Contribution Function-weighted mean of the temperature, such as in
Landi \& Landini (1997).

%

\subsection{Pitfalls of the diagnostic techniques}
\label{pitfalls}

Both the inversion and iterative DEM diagnostic methods rely on three main 
assumptions: 1) the interpolation
with a polynomial or a spline function assumes implicitly that the plasma is 
multithermal and may oversmooth the results; 2) the measured DEM depends on 
the widths of the temperature bins chosen to group the lines; and 3) with only
few exceptions, $T_{eff}$ is arbitrarily chosen as $T_{max}$, regardless of the 
thermal structure of the plasma.

The first problem makes the DEM analysis unsuitable for determining whether 
a plasma is isothermal or multithermal, because the DEM analysis implicitly 
assumes the latter condition. Also, associating each spectral line to a 
predetermined temperature like $T_{max}$ is likely to bias the results 
against an isothermal solution: in fact, if the plasma has a very narrow 
temperature distribution, the temperature of the plasma may be very different 
from the $T_{max}$ of most lines 
in the data set. Associating each line to $T_{max}$ results in associating
a line to a temperature where there might not be any plasma at all and this
leads to an artificially multithermal DEM.

If the width of the temperature bin is chosen too large, variations of 
the DEM with temperature occurring in temperature intervals smaller than 
the bin width are averaged out so that the final DEM is oversmoothed. 
If on the contrary the bin width is too narrow, spurious effects due to 
problems in individual lines and photon noise can severely affect the solution.

Moreover, the EM and DEM techniques require the use of lines emitted by a 
large number of different ions whose ion fractions are non-negligible in 
widely different temperature ranges. The use of only few lines greatly
limits the diagnostic capabilities of these techniques, as demonstrated
by Judge (2010) and Landi \& Klimchuk (2010).

\subsection{The Markov Chain Monte Carlo DEM diagnostic technique}

The problems associated with DEM interpolation and with the definition 
of $T_{eff}$ do not affect the Markov-Chain Monte Carlo (MCMC) technique 
developed by Kashyap \& Drake (1998). This technique is based on a Bayesian 
statistical formalism that allows the determination of the most probable
DEM curve that reproduces the observed line intensities. The heart of this 
technique relies on the application of the Bayes theorem to line intensities, 
stating that the probability $P{\left({X,F}\right)}$ of obtaining a set 
of observed line intensities $F={\left({F_1,F_2,...,F_n}\right)}$ from a DEM
characterized by a set of parameters $X={\left({X_1,X_2,...,X_m}\right)}$ 
is given by

\begin{equation}
P{\left({X,F}\right)}=P{\left({X}\right)}{{\Pi_{i=1,n}P{\left({X,F_i}\right)}}\over{P{\left({F}\right)}}}
\label{chap5:markow-1}
\end{equation}

\noindent
where $P{\left({F}\right)}$ is a normalization factor, $P{\left({X}\right)}$ is an
{\it a priori} probability of the set of parameters $X$, and $P{\left({X,F_i}\right)}$
is the probability of obtaining the observed intensity $F_i$ with the set of parameters $X$
and has been defined by Kashyap \& Drake (1998) as

\begin{equation}
P{\left({X,F_i}\right)} = e^{-{{\left({{\left({F_i-F_i^{th}}\right)}/{\sqrt{2}\sigma_i}}\right)}^2}}
\label{chap5:markow-2}
\end{equation}

\noindent
where $F_i^{th}$ is the intensity of line $i$ calculated with the DEM described 
by the set of parameters $X$, and $\sigma_i$ is the uncertainty of the observed 
intensity $F_i$. This method is implemented by choosing a grid of $N$ temperature
bins and assuming that within each bin the plasma is isothermal and can be 
described by an emission measure value $EM_i$. Once the bin size is chosen,
the set of parameters that describe the DEM is thus $X={\left({EM_1,...,EM_N}\right)}$.

In order to determine the set of parameters $X$ that provide the maximum
probability $P{\left({X,F}\right)}$, Kashyap \& Drake (1998) adopted a 
Markov-Chain Monte Carlo approach. In this approach, the set of parameters 
$X$ that describes an initial trial DEM is varied step by step; in each 
step only one of the parameters in the set $X$ (i.e. one $EM_i$ value only) 
is varied and the others are left unchanged; the change introduced to the 
varied parameter only depends on the set of parameters $X$ of the previous step. 
The new set of parameters $X'$ has a different probability $P{\left({X',F}\right)}$
from the previous one, and is accepted or rejected according to the Metropolis
algorithm (Metropolis \etal 1953), based on the change of probability $P$. This 
algorithm consists of generating a random number $u$, such that $0\leq u < 1$, 
and a function $A{\left({X,X'}\right)}$ defined as

\begin{equation}
A{\left({X,X'}\right)} = \min{\left[{1,{{P{\left({X',F}\right)}}\over{P{\left({X,F}\right)}}}}\right]}
\label{chap5:markow-3}
\end{equation}

\noindent
The new set of parameters is accepted if $u<A{\left({X,X'}\right)}$, and rejected
otherwise. In this way, not only is a new set of parameters $X'$ with greater
probability than the previous one always accepted (since $A{\left({X,X'}\right)}=1$),
but also a new set with a smaller probability has a finite chance of being accepted. 
This latter property helps finding the best distribution of parameters $X$ by moving 
the solution out of local maxima. When the system has found the best solution, the 
distribution of the $EM$ values in each bin found in all steps is used to determine 
the confidence interval at each temperature, thus providing an estimate of the 
uncertainty of the DEM at all temperatures.

This method dispenses from two of the assumptions common to the other two methods.
First, it is not affected by the choice of $T_{eff}$, since each line contributes to 
determine the solution in the entire temperature range where its $G{\left({T}\right)}$ 
function is defined. Also, no interpolation of the solution is required and no assumption
on the smoothness of the solution is made. However, the choice of the bin width is crucial 
for this technique as it is for the iterative and inversion ones, and it is necessary to 
assess its effect.

\subsection{Methodology of the tests}

In order to test the ability of the MCMC method to reconstruct the DEM, we 
apply this technique to line intensities calculated using known thermal
distributions and version 6.0.1 of
the CHIANTI database (Dere \etal 1997, 2009) using the Bryans \etal (2009) ion
fractions; thus, uncertainties 
due to the atomic physics do not affect our tests. We assumed a density value of
$\log N_e=9.0$ ($N_e$ in cm$^{-3}$), typical of moderately active region plasmas.
In order to understand how the MCMC diagnostic technique is affected by errors
in the data (both in the atomic data used for emissivity calculations as well as 
observational uncertainties), for each of the synthetic spectra we used we generated 
five additional datasets; in each additional dataset we have added
random errors of up to 20\%, i.e. we have randomized each line intensity
within a 20\% range around the nominal value. Thus, each thermal 
distribution we have considered will be reconstructed using first the 
original synthetic line intensities with no errors added, and then each of 
the five different sets of intensities with random errors. With five 
different ``noisy'' datasets, the stability of the solution against errors 
in the data can be evaluated. As a further test on the effects due to atomic
physics only, we repeated the MCMC determinations using the synthetic intensities
calculated with version 2 of the CHIANTI database (Landi \etal 1999) and the ion
fractions of Mazzotta \etal (1998). The atomic data and ion fractions in the two 
CHIANTI versions are very different, so that the use of CHIANTI V.2 represents
a rather bad case of uncertainties in atomic parameters.

\begin{figure}
\includegraphics[width=8.0cm,height=8.0cm,angle=90]{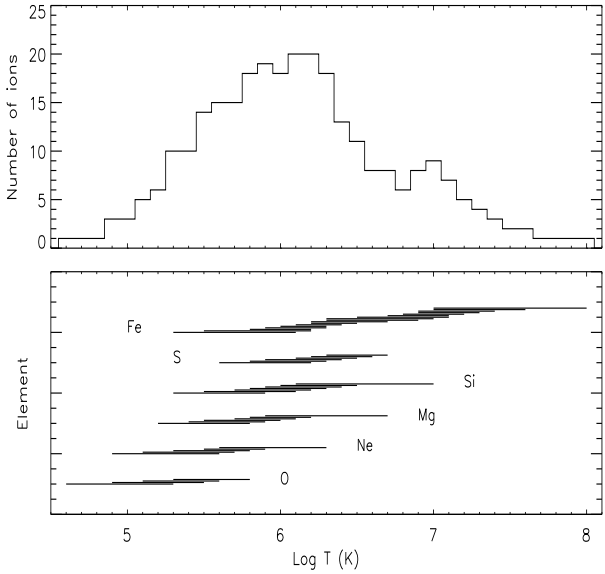}
\caption{\label{temp_coverage} Temperature coverage of the ions listed in
Table~\ref{lines}. For each ion only the temperature interval where the
fractional abundance is larger than 0.01 has been considered. Ion abundances
are from Bryans \etal (2009). {\bf Top:} number of ions emitting in each 
temperature bin (bin width $W=0.1$). {\bf Bottom:} Temperature range covered 
by the ions of each element.}
\end{figure}

We considered the particularly favorable situation of many lines available for 
diagnostics: we used lines from 45 different ions, listed in 
Table~\ref{lines}, that allow to sample very finely the temperature interval 
$4.6 \leq \log T \leq 8.0$, as shown by Figure~\ref{temp_coverage}. However, 
above $\log T=7.0$ and below $\log T=5.5$ the number of available ions 
decreases, and thus the MCMC reconstruction is expected to be less accurate.

\begin{table*}
\caption{Lines used in the present work.}
\label{lines}
\begin{center}
\begin{tabular}{lrrrr|lrrrr}
Ion & Wvl.(\AA) & $\log T_{max}$ & G${\left({T_{max}}\right)}$ & & Ion & Wvl.(\AA) & $\log T_{max}$ & G${\left({T_{max}}\right)}$ & \\
\hline
 & & & \\
\ion[O iii]   &   702.900 & 4.90 & 1.13e-13 &         & \ion[S ix]     &   871.726 & 6.05 & 1.97e-16 &         \\
\ion[O iv]    &   787.710 & 5.17 & 1.00e-12 &         & \ion[S x]      &   264.231 & 6.18 & 7.31e-15 & $\star$ \\
\ion[O v]     &   629.732 & 5.37 & 7.60e-12 & $\star$ & \ion[S xi]     &   281.402 & 6.28 & 2.45e-15 & $\star$ \\
\ion[O vi]    &  1031.914 & 5.48 & 2.91e-12 & $\star$ & \ion[S xii]    &   288.421 & 6.35 & 4.76e-15 & $\star$ \\
\ion[Ne iv]   &   543.887 & 5.21 & 6.53e-14 &         & \ion[S xiii]   &   256.685 & 6.42 & 1.75e-14 & $\star$ \\
\ion[Ne v]    &   572.336 & 5.42 & 1.00e-13 &         & \ion[Fe viii]  &   185.213 & 5.62 & 9.55e-14 & $\star$ \\
\ion[Ne vi]   &   558.685 & 5.59 & 1.02e-13 &         & \ion[Fe ix]    &   171.073 & 5.87 & 5.24e-13 & $\star$ \\
\ion[Ne vii]  &   465.220 & 5.71 & 5.47e-13 &         & \ion[Fe x]     &   177.240 & 6.04 & 1.12e-13 & $\star$ \\
\ion[Ne viii] &   770.410 & 5.80 & 2.54e-13 &         & \ion[Fe xi]    &   182.169 & 6.13 & 2.87e-14 & $\star$ \\
\ion[Mg v]    &   276.579 & 5.45 & 2.51e-14 & $\star$ & \ion[Fe xii]   &   193.509 & 6.19 & 9.84e-14 & $\star$ \\
\ion[Mg vi]   &   270.391 & 5.63 & 2.11e-14 & $\star$ & \ion[Fe xiii]  &   202.044 & 6.25 & 6.58e-14 & $\star$ \\
\ion[Mg vii]  &   276.154 & 5.78 & 9.33e-15 & $\star$ & \ion[Fe xiv]   &   264.790 & 6.29 & 5.79e-14 & $\star$ \\
\ion[Mg viii] &   782.364 & 5.90 & 6.18e-15 &         & \ion[Fe xv]    &   284.163 & 6.34 & 3.18e-13 & $\star$ \\
\ion[Mg ix]   &   706.036 & 5.99 & 1.95e-14 &         & \ion[Fe xvi]   &   262.976 & 6.43 & 1.10e-14 & $\star$ \\
\ion[Mg x]    &   609.794 & 6.07 & 1.54e-13 &         & \ion[Fe xvii]  &   204.665 & 6.61 & 9.20e-16 & $\star$ \\
\ion[Si vi]   &   246.003 & 5.61 & 2.91e-14 & $\star$ & \ion[Fe xviii] &   974.860 & 6.86 & 7.91e-15 &         \\
\ion[Si vii]  &   275.361 & 5.78 & 6.15e-14 & $\star$ & \ion[Fe xix]   &  1118.057 & 6.95 & 5.14e-15 &         \\
\ion[Si viii] &   276.850 & 5.93 & 1.13e-14 &         & \ion[Fe xx]    &   132.840 & 7.01 & 7.55e-15 &         \\
\ion[Si ix]   &   258.082 & 6.05 & 3.18e-15 & $\star$ & \ion[Fe xxi]   &  1354.067 & 7.06 & 1.27e-14 &         \\
\ion[Si x]    &   258.371 & 6.15 & 3.78e-14 & $\star$ & \ion[Fe xxii]  &   845.571 & 7.11 & 6.10e-15 &         \\
\ion[Si xi]   &   580.920 & 6.22 & 9.08e-15 &         & \ion[Fe xxiii] &   263.766 & 7.17 & 1.73e-15 &         \\
\ion[Si xii]  &   520.666 & 6.30 & 4.48e-14 &         & \ion[Fe xxiv]  &   192.029 & 7.26 & 1.17e-14 & $\star$ \\
\ion[S viii]  &   198.554 & 5.91 & 3.93e-15 &         &                &           &      &          &         \\
 & & & \\
\hline
\end{tabular}
\tablefoot{$T_{max}$: temperature of maximum ion abundance. G${\left({T_{max}}\right)}$: 
line emissivity at $T_{max}$. Stars indicate the ions whose lines can be observed
by the Hinode/EIS spectrometer (although EIS lines may be different from those listed here).}
\end{center}
\end{table*}

The arbitrary width of the temperature bin has been varied, in order to 
determine the optimal value that allows us to best reconstruct the initial 
thermal distribution and its temperature dependence while limiting 
both the noise and variability due to a too small bin width and 
the oversmoothing due to a too large width. We run our tests using a 
grid of width $W$ values: $W=0.01,0.02,0.05,0.1,0.2$.

We considered several types of thermal structure:

\begin{enumerate}
\item Single isothermal plasma;
\item Two isothermal plasmas, with variable peak temperature separation and variable relative peak amplitude;
\item A single Gaussian distribution of variable width;
\end{enumerate}

\noindent
These thermal distributions are aimed at determining 1) to what extent 
and with what precision is the MCMC diagnostic technique able to detect an isothermal
plasma; 2) how capable is it at separating two isothermal components close in
temperature; 3) to what extent is it able to discriminate between an isothermal 
component and a multithermal plasma characterized by a Gaussian distribution; and 
4) how do random errors and atomic physics uncertainties affect all the above 
results. 

\subsection{Uncertainties of the results}

One of the main advantages of the MCMC technique is the ability to provide an
estimate of the confidence level of the solution. To determine it, we used the
following procedure. Once the best solution is found, we have run additional
50,000 runs of the Markov Chain procedure allowing each bin to randomly change
by factors up to 6 orders of magnitude. For each change, we calculated the 
$\chi^2$ of the solution and its probability, defined by Equation~\ref{chap5:markow-2}.
We arbitrarily retained only the modified EM distributions with a probability of 
at least 0.1$\times P_{best}$, where $P_{best}$ is the probability  of the best 
solution. Then, for each temperature bin we have calculated the average $\log EM$ 
value and its variance, and taken the value of the latter as an indication of the 
uncertainty on the EM value of that bin in the best solution. 

\subsection{Evaluation of the results}
\label{evaluation}

The resulting EM${\left({T}\right)}$ curve is compared to the EM distribution
used to calculate the synthetic spectra. The parameters we used to assess the 
quality of a reconstruction are shown in Table~\ref{metrics}. In this Table, 
we report the cases where the plasma is isothermal with $\log T=6.0$, or made 
as the sum of two isothermal components with temperatures $\log T_1=6.0$ and 
$\log T_2=\log T_1+ \Delta \log T$. The temperature distance of the two 
isothermal components in the multithermal case is chosen to be $\Delta \log 
T=0.10, 0.15, 0.20$. A successful fit to the original DEM curve is evaluated 
using the following criteria: 

\begin{enumerate}
\item Small $\chi^2$;
\item the DEM(T) curve should have a single peak (or two peaks for the multithermal 
case) at the temperatures $\log T_1$ and $\log T_2$; 
\item there should be no spurious components far from $T_1$ and $T_2$; 
\item the peak should be unique, i.e. there should not be spurious peaks 
anywhere near $T_1$ and $T_2$ (noise in the solution);
\item the width of the peak (taken as the temperature range around $T_1$ and 
$T_2$ where the EM(T) value is 1/10 of the peak value or larger) should be 
as small as the temperature bin $W$. 
\end{enumerate}


\begin{table}
\begin{tabular}{l|cccc}
               & W=0.01       & W=0.02       & W=0.05       & W=0.1       \\
\hline
 & & & & \\
 & \multicolumn{4}{c}{Isothermal plasma - no random error} \\
 & & & & \\
$\chi^2$       & 0.013        & 0.037        & 0.0018       & 0.00063      \\
Spurious comp. & Y            & N            & N            & N            \\
Single peak    & N            & N            & Y            & Y            \\
Noise          & Y            & N            & N            & N            \\
EM(T) width    & 0.03         & 0.04         & 0.05         & 0.10         \\
$\log T_1$     & 5.99-6.03    & 5.98-6.02    & 6.00         & 6.00         \\
 & & & & \\
\hline
 & & & & \\
 & \multicolumn{4}{c}{Multithermal plasma - no random error} \\
 & & & & \\
 & \multicolumn{4}{c}{\underline{$\Delta \log T=0.10$}} \\
 & & & & \\
$\chi^2$       & 0.0062       & 0.0085       & 0.0058       & 0.0069       \\
Spurious comp. & Y            & N            & N            & N            \\
Two peaks      & N            & N            & Y            & Y            \\
Noise          & Y            & Y            & N            & N            \\
EM(T) width    & 0.03         & 0.06         & 0.05         & 0.10         \\
$\log T_1$     & 5.99-6.02    & 5.98-6.04    & 6.00         & 6.00         \\
$\log T_2$     & 6.10-6.11    & 6.10         & 6.10         & 6.10         \\
 & & & & \\
 & \multicolumn{4}{c}{\underline{$\Delta \log T=0.15$}} \\
 & & & & \\
$\chi^2$       & 0.0013       & 0.0035       & 0.00087      & 0.23         \\
Spurious comp. & Y            & Y            & N            & N            \\
Two peaks      & N            & N            & Y            & Y            \\
Noise          & Y            & Y            & N            & N            \\
EM(T) width    & 0.03         & 0.06         & 0.05         & 0.20         \\
$\log T_1$     & 5.99-6.02    & 5.98-6.04    & 6.00         & 6.00-6.20    \\
$\log T_2$     & 6.15         & 6.14-6.16    & 6.15         & 6.00-6.20    \\
 & & & & \\
 & \multicolumn{4}{c}{\underline{$\Delta \log T=0.20$}} \\
 & & & & \\
$\chi^2$       & 0.0053       & 0.0078       & 0.0017       & 0.00055      \\
Spurious comp. & Y            & N            & N            & N            \\
Two peaks      & N            & N            & Y            & Y            \\
Noise          & Y            & Y            & N            & N            \\
EM(T) width    & 0.04         & 0.04         & 0.05         & 0.10         \\
$\log T_1$     & 5.99-6.02    & 5.98-6.02    & 6.00         & 6.00         \\
$\log T_2$     & 6.18-6.22    & 6.18-6.22    & 6.20         & 6.20         \\
 & & & & \\
\hline
\end{tabular}
\caption{\label{metrics} Performance of the MCMC technique when applied to
isothermal plasmas, and to multithermal plasmas composed of two different
isothermal components. See Section~\ref{evaluation} for details.}
\end{table}

\section{Results}
\label{results}

\subsection{Isothermal plasmas}
\label{isot-comp}

\begin{figure}
\includegraphics[width=6.0cm,height=8.0cm,angle=90]{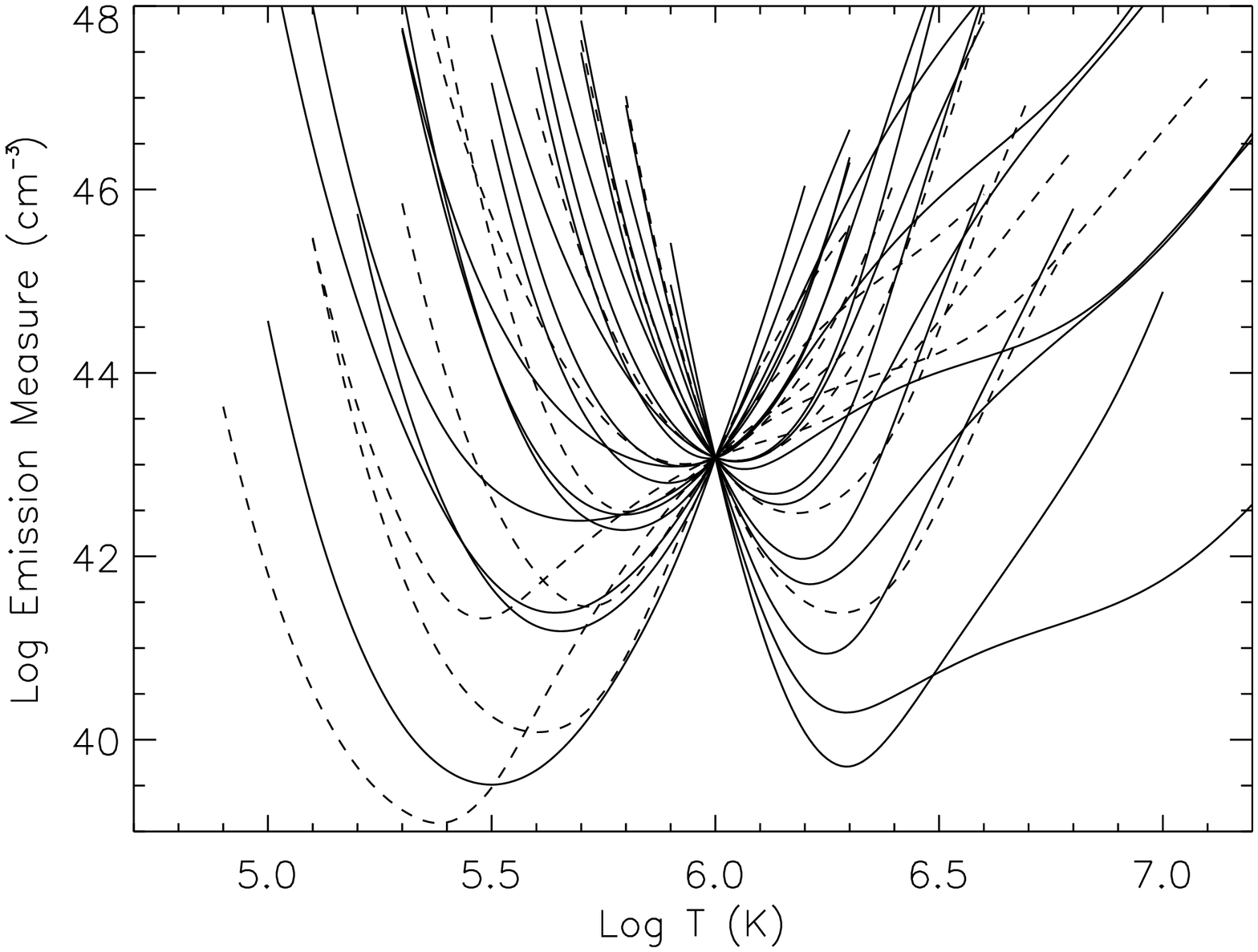}
\caption{\label{1-isot-1} EM loci technique applied to the single isothermal plasma, without random
errors.}
\includegraphics[width=7.0cm,height=8.0cm,angle=90]{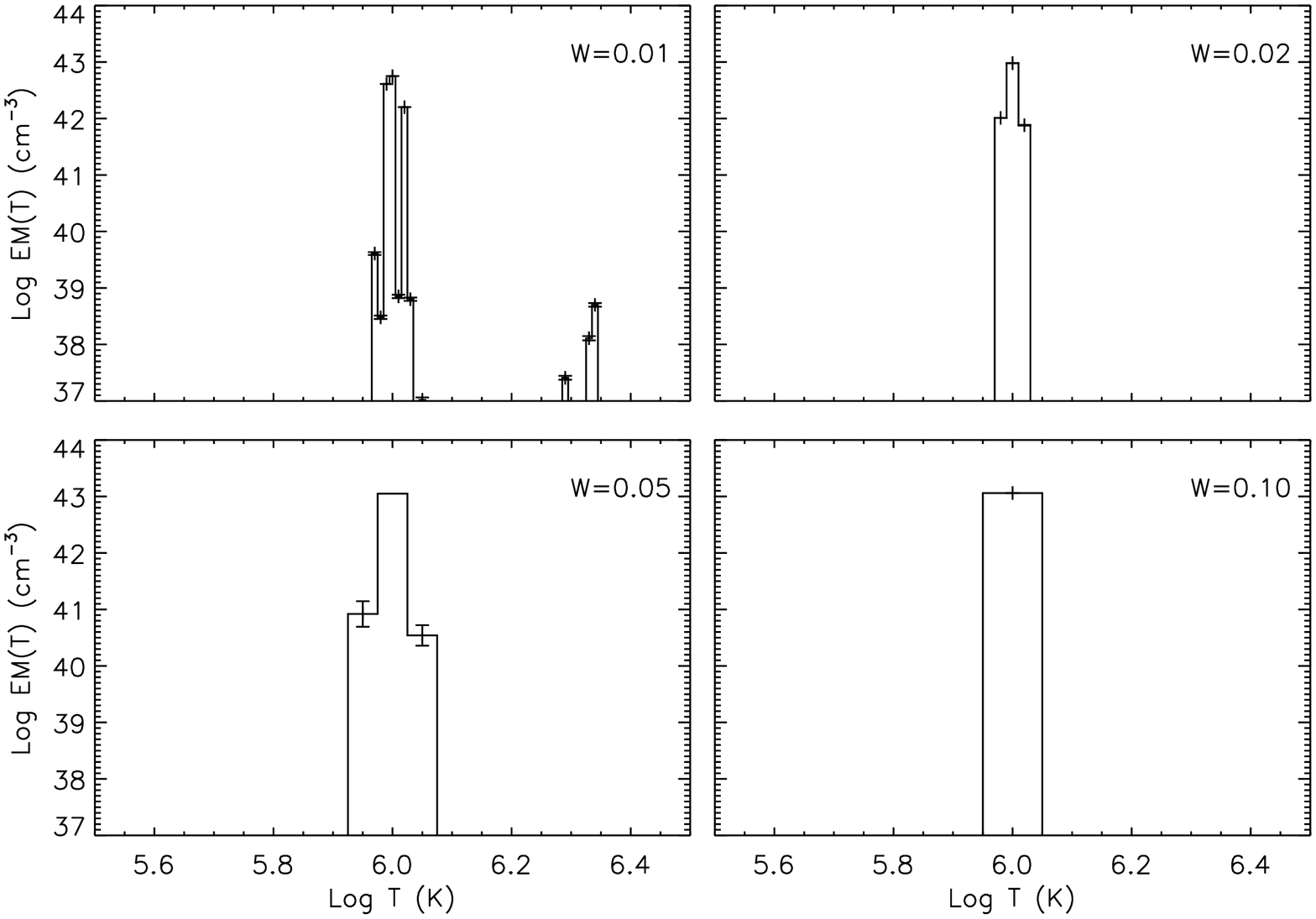}
\caption{\label{1-isot-3} $EM(T)$ reconstruction of a single isothermal plasma, without random errors.
The width of the temperature bin changes from $W=0.01$ to 0.10.}
\includegraphics[width=7.0cm,height=8.0cm,angle=90]{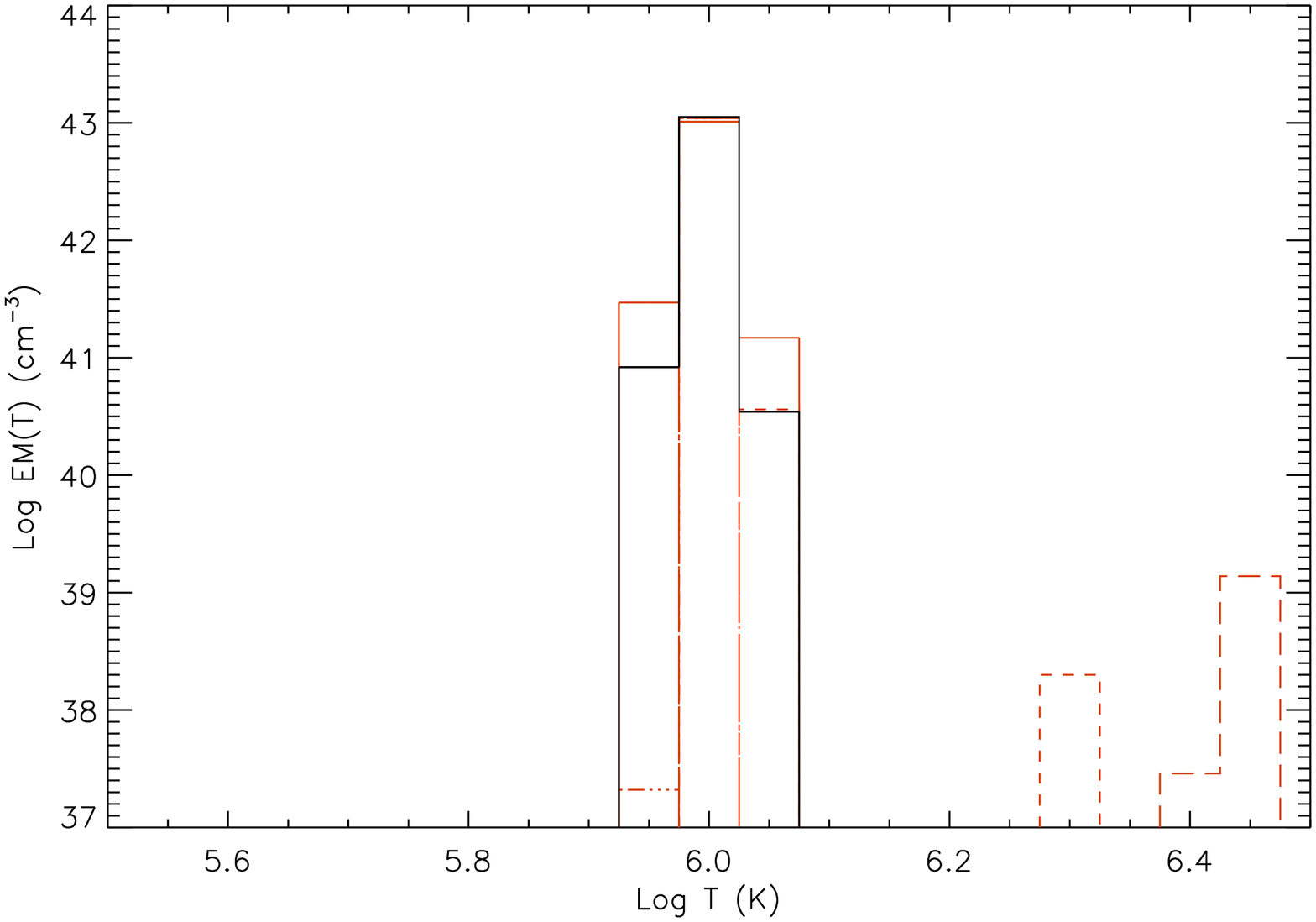}
\caption{\label{1-isot-2} $EM(T)$ reconstruction of a single isothermal plasma, with 
random errors within 20\% added to line intensities (see text for details). The width 
of the temperature bin is $W=0.05$. All five ``noisy'' datasets are shown.}
\end{figure}

The isothermal plasma was chosen to have $\log T=6.0$ (in K) and $\log EM=43.0$ (in 
cm$^{-3}$). This temperature was chosen as it allowed us to include lines routinely 
observed by available spectrometers such as SOHO/CDS, SOHO/SUMER, and Hinode/EIS. 
Figure~\ref{1-isot-1} shows the EM loci technique applied to the calculated intensities,
without random errors included: the original temperature and EM values are recovered
without problems.
Figure~\ref{1-isot-3} shows the MCMC reconstruction obtained without random errors 
and varying the bin size $W$. Results are summarized in Table~\ref{metrics}. In all 
cases the MCMC technique is able to provide an 
isolated peak several order of magnitude larger than the background. In the case 
of bin width $W=0.01$ the shape of the peak is irregular and one additional 
satellite peak is present at higher temperatures: this feature is an artifact 
and it disappears when the bin width increases.
In all cases the peak maintains a non-negligible width $\Delta{\left({\log T}\right)} 
\approx 0.03-0.05$ that is approximately constant as the bin width $W$ increases. 
We interpret this as the ultimate resolution capability of the MCMC technique to 
resolve a purely isothermal plasma: the result is indistinguishable from a 
multithermal plasma with a very narrow width. Even if the $\chi^2$ of the
solution with W=0.1 is the lowest, the solution with W=0.05 is to be preferred
because the bin width is closer to the apparently intrinsic resolving power of 
the MCMC technique. In the case with $W=0.05$ the two
bins at the side of the peak show EM values of $\approx$1\% of the peak value, 
and might be taken as real. Therefore, care should be taken in interpreting such 
features.

The solutions provided by the MCMC technique always have a background at very low EM 
values at all temperatures ($\log EM \simeq 33-37$). This background represents the 
maximum values of the EM 
at all temperatures outside the isothermal peak that provide contributions to the 
$\chi^2$ smaller than the computer precision. However, this background has no
physical significance.

Figure~\ref{1-isot-2} shows how the results change when the random errors are
added to the calculated intensities. The plasma EM obtained with $W=0.05$ is displayed 
as a full black line, and the EM curves obtained by the five different ``noisy'' datasets 
are shown in red. The main peak remains unaltered, and since the wings,
always present, are much lower than the peak value the technique is still able to 
recover the single peak. However, the background noise now is more significant and 
secondary peaks, much lower than the one at $\log T=6.0$, are sometimes present such 
as those at $\log T=6.30$ and 6.45.

\subsection{Two-component plasmas}

\begin{figure*}
\centerline{\includegraphics[width=17.0cm,height=15.0cm,angle=90]{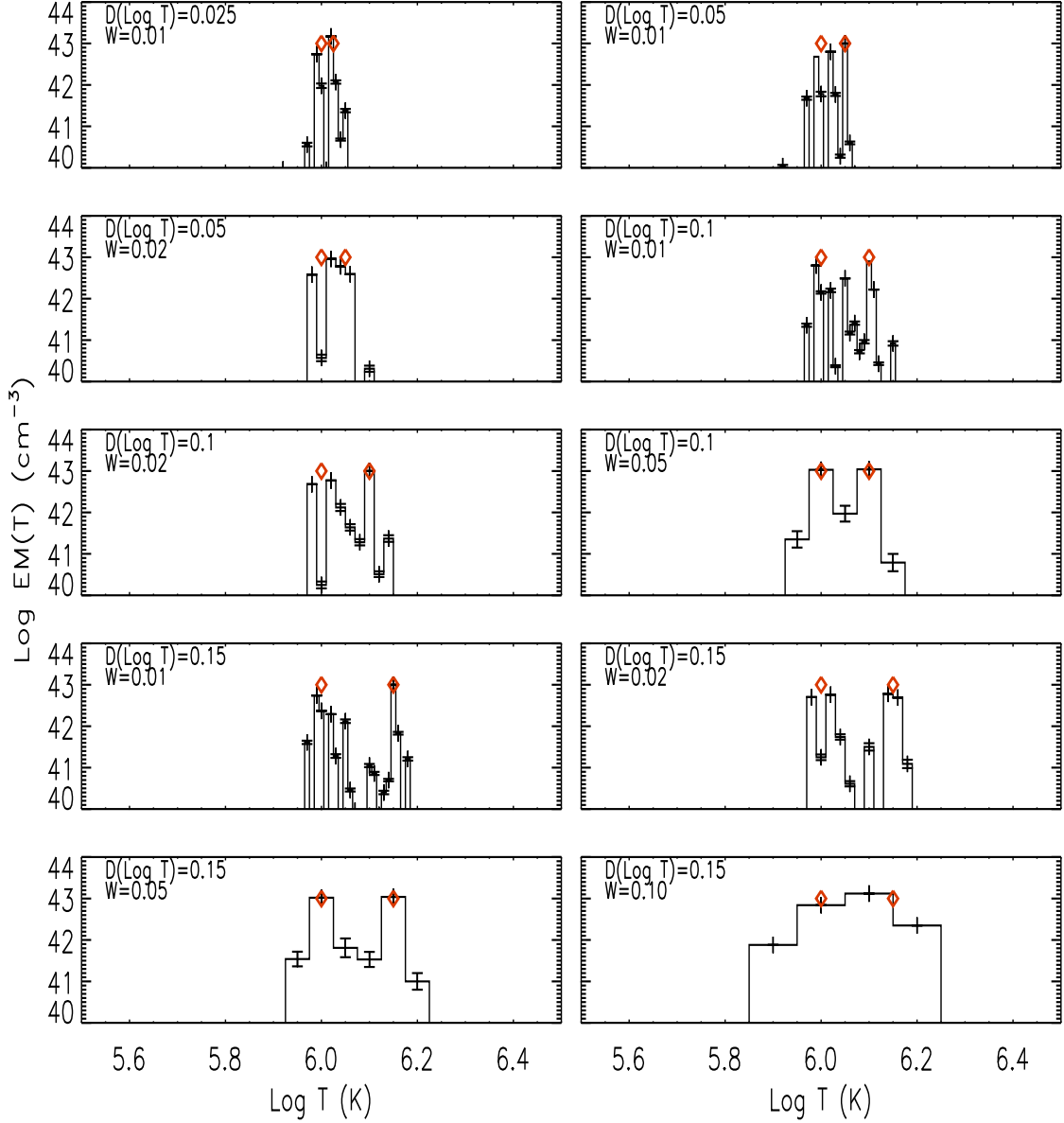}}
\caption{\label{2-isot-1} $EM(T)$ reconstruction of a a plasma made of two isothermal
components whose peak temperatures are separated by $D(\log T)$, which is varied
from 0.025 to 0.15. The bin width $W$, reported in each panel, is chosen to be always 
smaller than the peak separation $D(\log T)$.}
\end{figure*}

Figure~\ref{2-isot-1} displays the results of the reconstruction of the $EM(T)$ curve in 
the case of two isothermal components with the same EM ($\log 43.0$ each, in cm$^{-3}$,
shown as red diamonds) 
and with variable temperature separation $\Delta \log T$ and no random errors. The bin 
width $W$ in each reconstruction was kept smaller than the temperature separation, and 
when possible we experimented with increasing $W$ to determine how it affected the
ability of the technique to separate the two components. We also studied $\Delta \log 
T=0.20$ and 0.25, but the results are no different than those obtained with $\Delta \log 
T=0.15$. Table ~\ref{metrics} reports the results for the reconstructions obtained
with $\Delta \log T=0.10, 0.15$ and 0.20.

Figure~\ref{2-isot-1} shows that even when random errors are absent the MCMC technique 
is unable to resolve the two components when their temperature separation is smaller 
than $\Delta \log T=0.10$, and provides a single, very noisy peak with a larger width. 
Also, small values of the bin width $W$ worsen the noise in the solution, further 
preventing the separation of the two components. At $\Delta \log T=0.10$ a 
complete separation of the two peaks can be achieved only when $W=0.05$, because 
noise confuses the result for smaller bin widths and the presence of high EM values 
in the only bin between the peaks prevents definitive conclusions about separate 
components. Only when $\Delta \log T=0.15$ the two components can be fully and 
convincingly resolved with $W=0.05$. Noise in the definition of each peak is 
significant at any value of $\Delta Log T$ for $W<0.05$; the bin $W=0.1$ allows us 
to resolve the two peaks only when $\Delta \log T=0.20$ and causes trouble if the 
separation is $\Delta \log T=0.15$, providing a large $\chi^2$.

\begin{figure}
\includegraphics[width=5.0cm,height=8.0cm,angle=90]{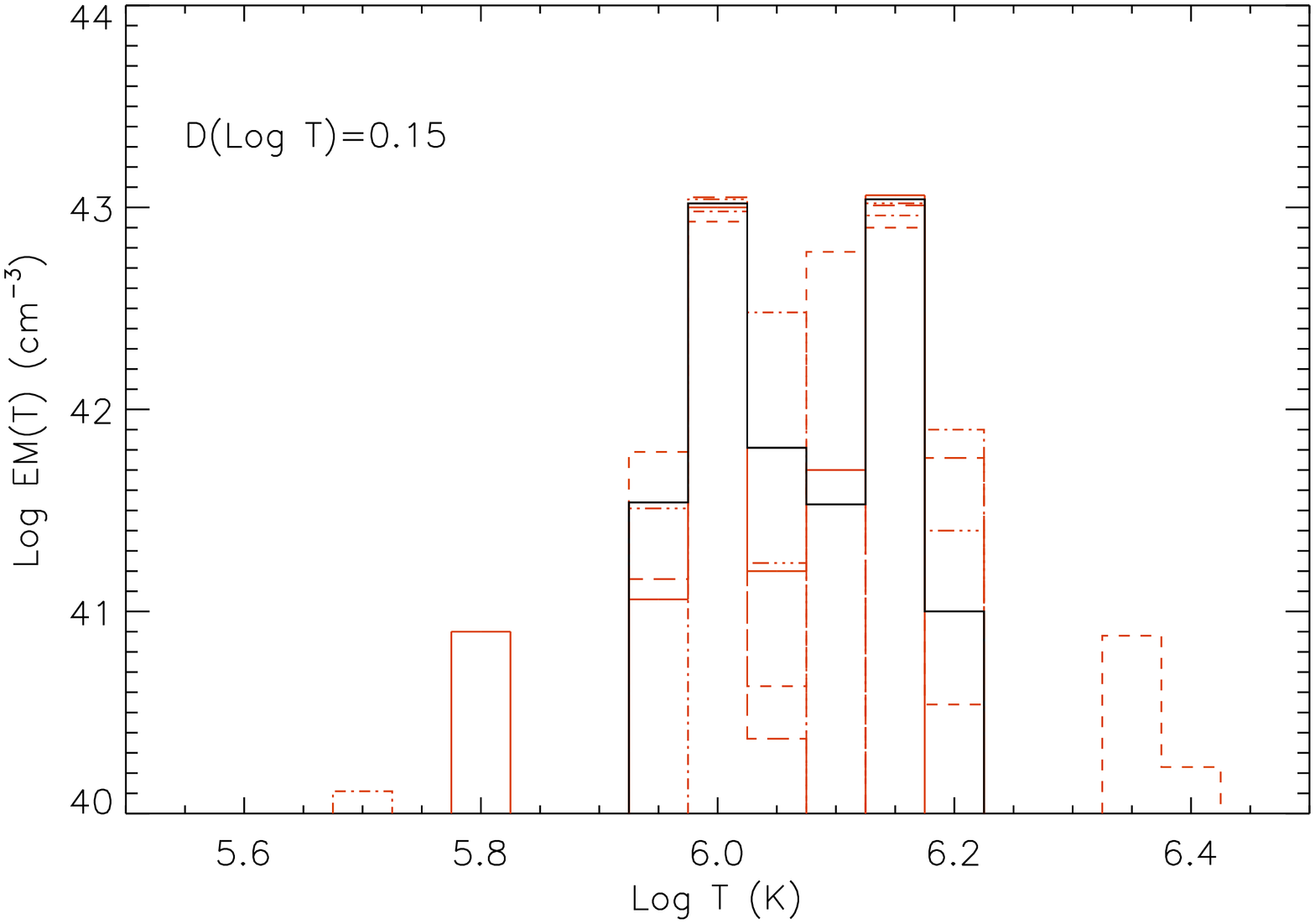}
\includegraphics[width=5.0cm,height=8.0cm,angle=90]{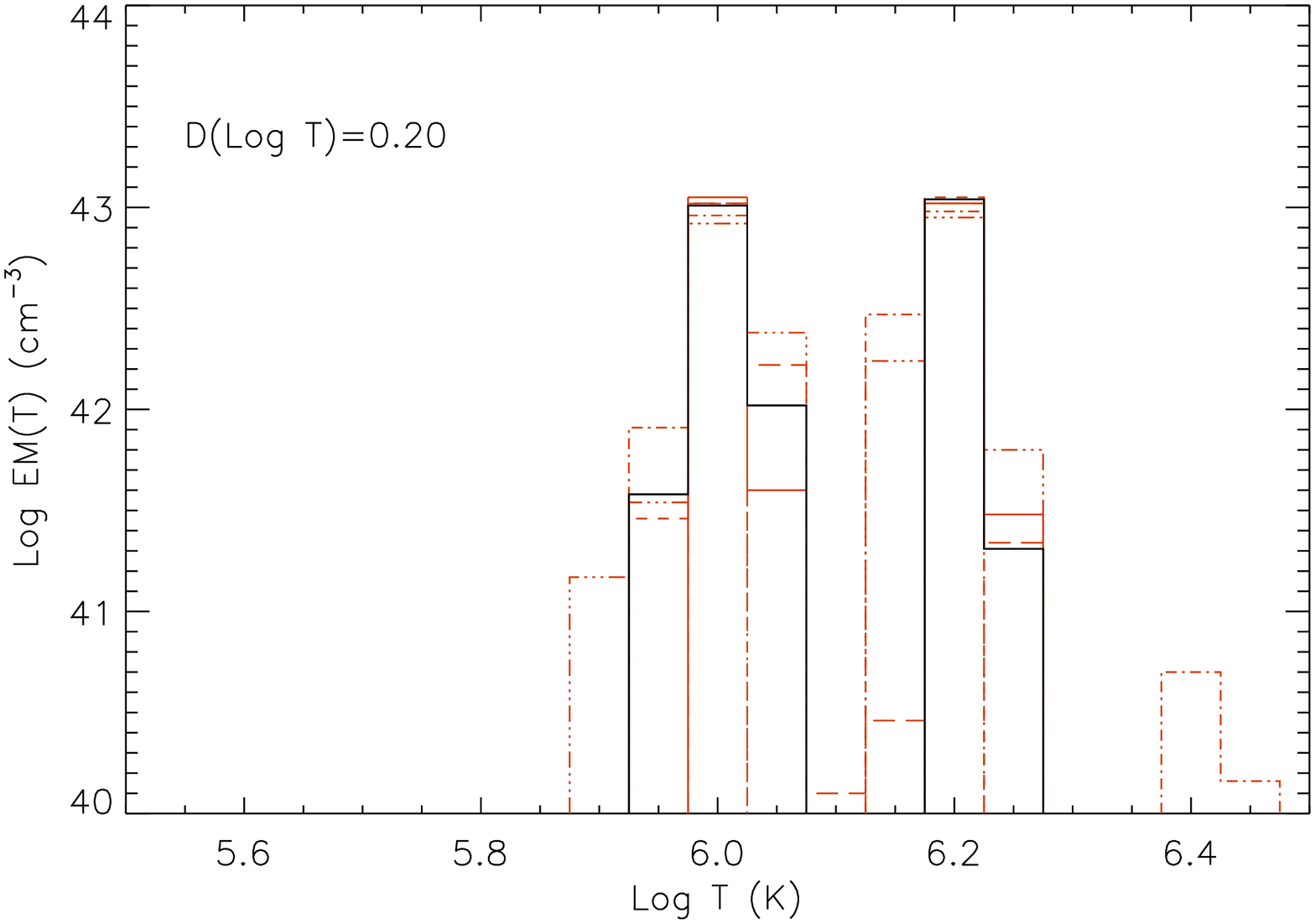}
\caption{\label{2-isot-2} $EM(T)$ reconstruction of a a plasma made of two isothermal
components whose peak temperatures are separated by $D(\log T)=0.15$ ({\bf top}) and
0.20 ({\bf bottom}). The bin width $W$ is 0.05. Line intensities have been added a random
errors within 20\% (see text for details). The reconstruction has been made without 
(black full curve) and with (red dashed curves) a random errors within 20\% to each 
line intensity.}
\end{figure}

The presence of random errors further limits the ability of the MCMC technique to
resolve the two components. Figure~\ref{2-isot-2} shows the MCMC reconstruction for 
$\Delta \log T=0.15$ and 0.20 in the presence of random errors, with $W=0.05$. The two
peaks are still visible but badly resolved when $\Delta \log T=0.15$, while they
are well separated in the other case. Thus, we conclude that $\Delta \log T=0.20$ 
is a reliable estimate of the smallest temperature distance between two
isothermal components which the MCMC technique can realistically resolve
when the lines listed in Table~\ref{lines} are available. If only a smaller number 
of lines can be used, the minimum $\Delta \log T$ value may be significantly larger. 
It is also important to note the spurious components that arise around the two 
isothermal components when the random errors are present (e.g. at $\log T=5.7, 5.8, 
6.35$ etc). Their peak values are significantly lower than the true components, but 
nevertheless they complicate the interpretation of the result.

\begin{figure}
\includegraphics[width=5.0cm,height=8.0cm,angle=90]{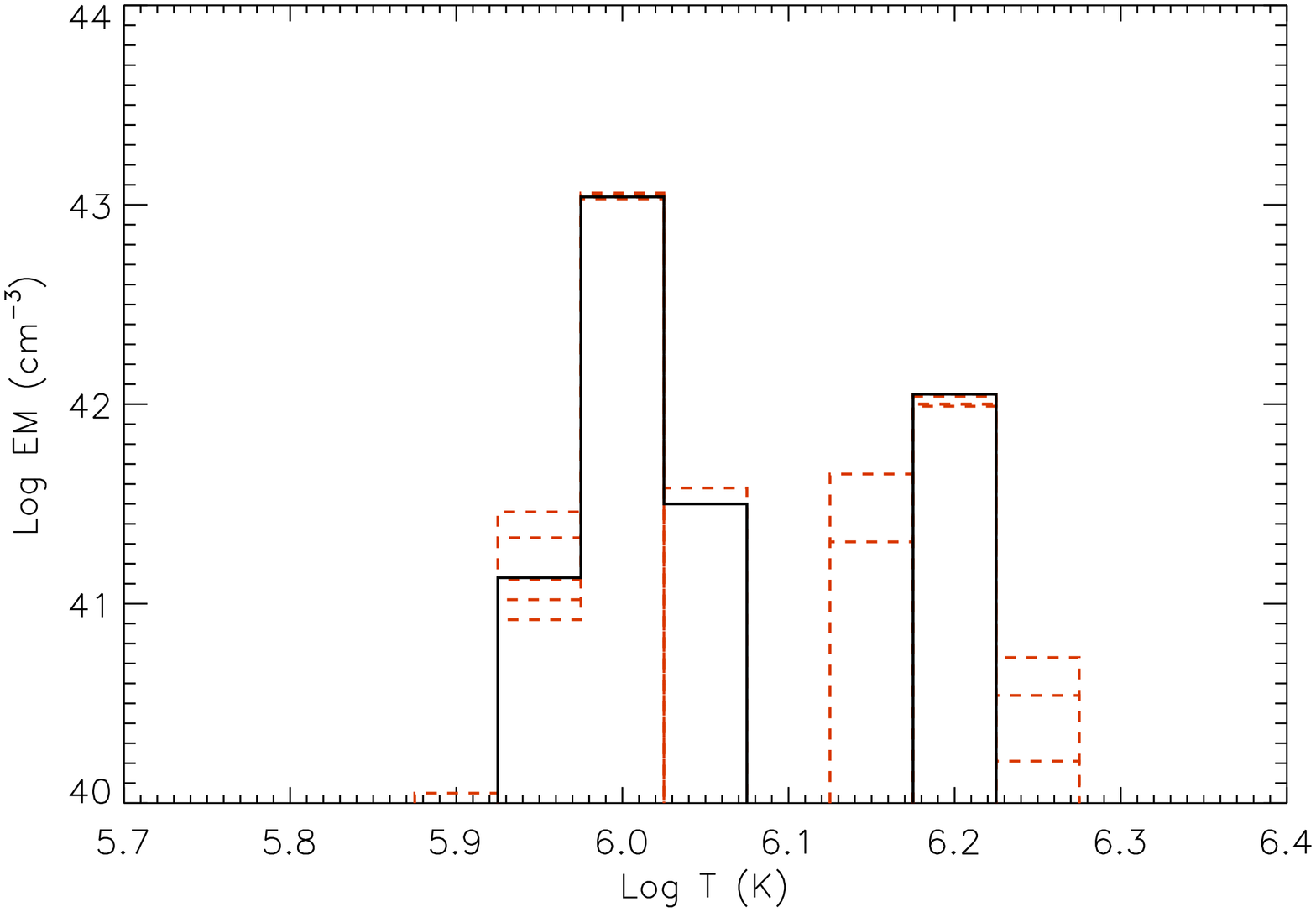}
\includegraphics[width=5.0cm,height=8.0cm,angle=90]{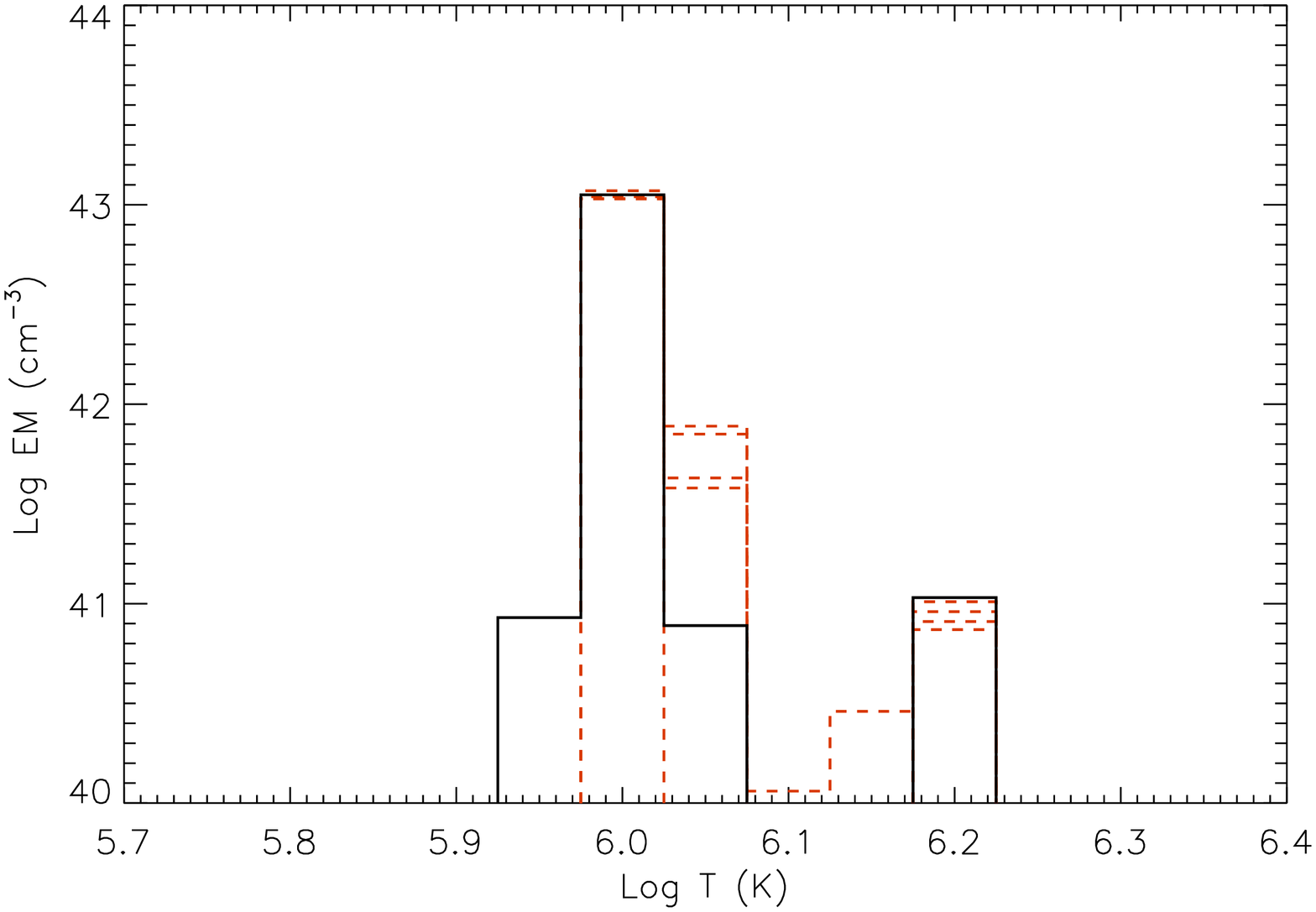}
\caption{\label{2-isot-var} $EM(T)$ reconstruction of a plasma made of two isothermal
components whose peak temperatures are $\log T=6.0$ and 6.2. The hotter peak EM is a 
factor 10 ({\bf top}) and 100 ({\bf bottom}) smaller than the colder one. The bin width 
$W$ is 0.05. The reconstruction has been made without (black full curve) and with (red 
dashed curves) a random errors within 20\% to each line intensity.}
\end{figure}

The MCMC technique maintains its ability to resolve double peaks even when their
relative size is much different. Tests have shown that when the peak separation is
$\Delta \log T=0.20$ the two peaks can still be resolved in the presence of random errors
even if one of them is a factor 10 or even 100 smaller than the other as shown in 
Figure~\ref{2-isot-var} where MCMC reconstructions of a plasma distribution with
two isothermal components separated by $\Delta \log T=0.20$ with different heights
are shown without (black full lines) and with (red dashed lines) the presence of
random errors. The peaks can be recognized because the peak EM values are 
very close in all reconstructions, while in the other temperature bins the 
EM values are scattered over a broad EM range.

\subsection{Gaussian DEM distribution}

\begin{figure*}
\includegraphics[width=11.0cm,height=15.0cm,angle=90]{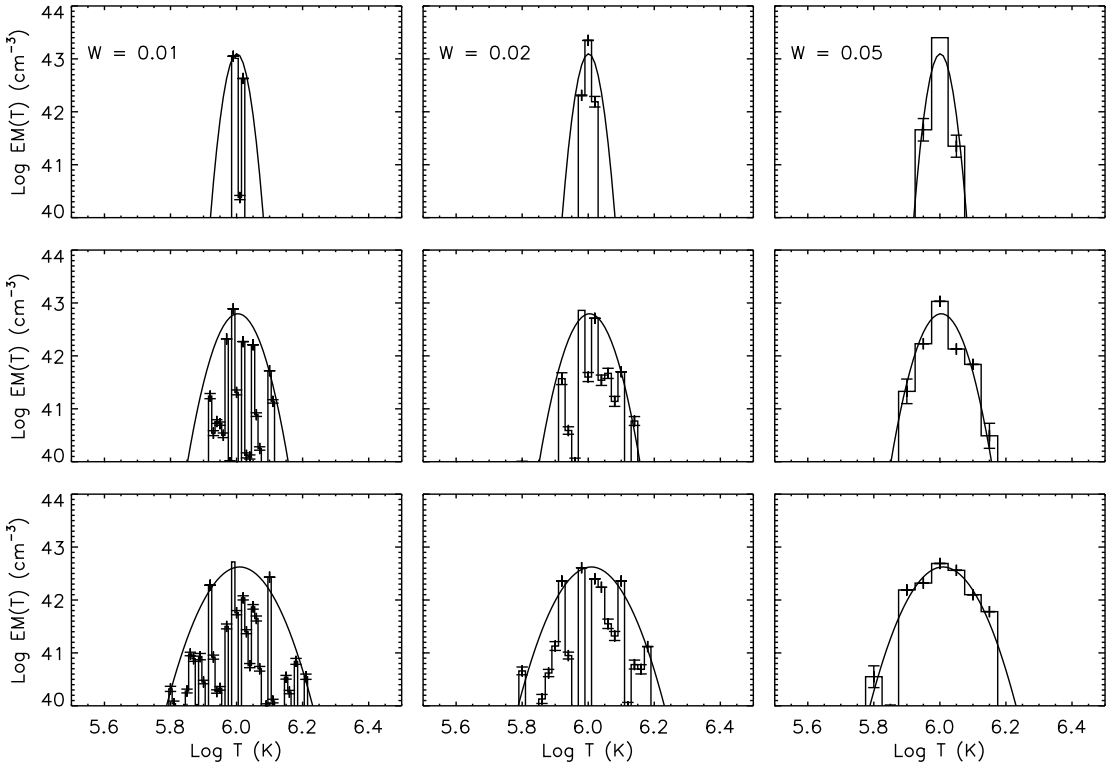}
\caption{\label{gaussian} MCMC reconstructions of Gaussian $EM(T)$ distributions.
Original curves are also shown.
The temperature bin width $W$ is 0.01, 0.02 and 0.05 (left, middle and center 
column, respectively). The Gaussian FWHM is 0.05 (top row), 0.10 (second row), 
and 0.15 (bottom row).}
\end{figure*}

The reconstruction of the Gaussian $EM(T)$ is shown in Figure~\ref{gaussian}. The 
Gaussian $EM(T)$ curves were defined with a variable FWHM (in $\log T$): 0.05, 0.10, 
0.15. The $EM(T)$ curves are also displayed in Figure~\ref{gaussian}, starting
from top (FWHM=0.05) to bottom (FWHM=0.15). The reconstruction was performed with 
different widths $W=0.01, 0.02, 0.05$. When $W<0.05$, the solutions are too noisy 
to be acceptable in all cases, and also provide a solution too narrow when FWHM=0.05.
When $W=0.05$, the reconstruction is smoother and follows more closely the original 
$EM(T)$ curve.

\begin{figure}
\includegraphics[width=8.0cm,height=8.0cm,angle=90]{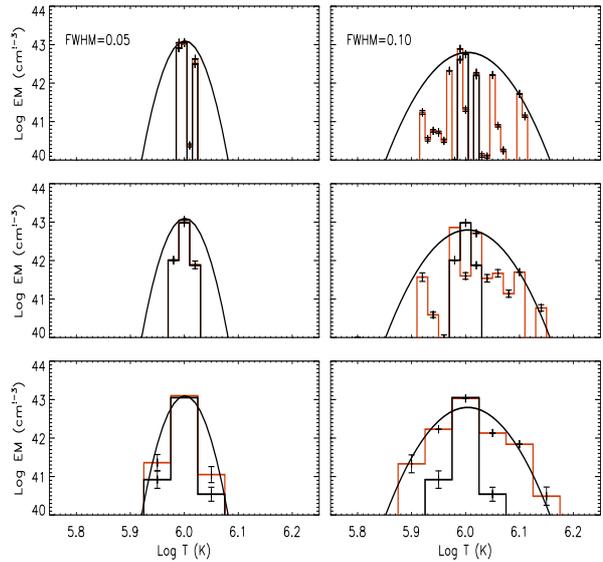}
\caption{\label{isot_vs_gauss} Comparison of MCMC reconstruction of a
single isothermal plasma (black) and a Gaussian $EM(T)$ curve (red).
The FWHM of the Gaussian is 0.05 (left column) and 0.10 (right column). 
The temperature bin $W$ is 0.01 (top row), 0.02 (middle row) and 0.05
(bottom row).}
\end{figure}

When FWHM=0.05, the solution provided by the MCMC technique is very similar to the one
provided in the single isothermal plasma component. Figure~\ref{isot_vs_gauss} compares
the isothermal results for the isothermal plasma component (black) and the Gaussian
$EM(T)$ (red): the two reconstructed curves are almost identical for FWHM=0.05, while
the Gaussian reconstruction is significantly larger than the isothermal one when 
FWHM=0.10. This means that the MCMC curve has problems distinguishing an
isothermal plasma from a Gaussian plasma when the width of the latter is very small.
This result means that the MCMC diagnostic technique is unable to discriminate between 
isothermal and narrow multithermal distributions. If we combine this limitation with 
the intrinsic width found in the single, isothermal plasma results, we can state that 
the MCMC technique is unable to unambiguously detect an isothermal plasma, and that 
is at best able to determine that a plasma distribution can have a $\log T$ 
width of 0.05 or smaller.

\subsection{Effects of a smaller number of ions}

The tests that we have carried out so far included lines from forty-five ions, much
more than those usually available in real-case scenarios of high-resolution observations
carried out with existing instruments like Hinode/EIS, SOHO/CDS, and SOHO/SUMER. These 
tests
show the intrinsic strengths and weaknesses of the MCMC technique. However, the number
of spectral lines and the amplitude of the temperature range they are formed in are also
limiting factors in the overall accuracy of DEM diagnostic techniques, as shown by
Judge (2010) and Landi \& Klimchuk (2010). The question is, how do the results change 
when a smaller number of ions is available.

To investigate this, we considered the most recent high-resolution spectrometer available
for studies of the solar atmosphere: Hinode/EIS (Culhane \etal 2007). We carried out the 
same tests described
in the previous sections using only lines from the ions that can be observed by EIS.
These are marked by asterisks in Table~\ref{lines}, and they are a total of 24 ions; 
we have omitted \ion[Si viii] and \ion[S viii] since their EIS lines are too heavily 
blended with other species to be used for DEM analysis. On the overall, the results 
that we have reached in the previous sections are confirmed.
Often, even smaller groups of ions than those available in the EIS wavelength range are used.
This is due to the fact that when EIS observations are designed and carried out, instrumental
or observational constraints usually make it necessary to transmit to the ground only portions
of the EIS spectral range, and to sacrifice several ions. A typical dataset is the one used
by Brooks \etal (2009), where the ions used were \ion[Si vii], \ion[Si x], \ion[S x],
and \ion[Fe viii-xv], for a total of 11 ions. Similar numbers are typical of most SOHO/CDS
and SOHO/SUMER observations as well. Results obtained from such a restricted dataset do not 
have significant differences when dealing with plasmas made of one or more isothermal components, 
and the MCMC technique is able to provide the same results as with larger datasets, provided
that the plasma temperature lies inside the range of formation of these ions. 

\subsection{Effects of inaccuracies in atomic data}

We test how systematic errors due to uncertainties in atomic physics affect the 
DEM diagnostics by creating an inconsistency between the generation and the reconstruction 
of the test DEM. In particular, we use an older version (V.2) of the CHIANTI atomic code to
calculate the contribution functions to be used for the DEM reconstruction.

\begin{figure}
\includegraphics[width=5.0cm,height=8.0cm,angle=90]{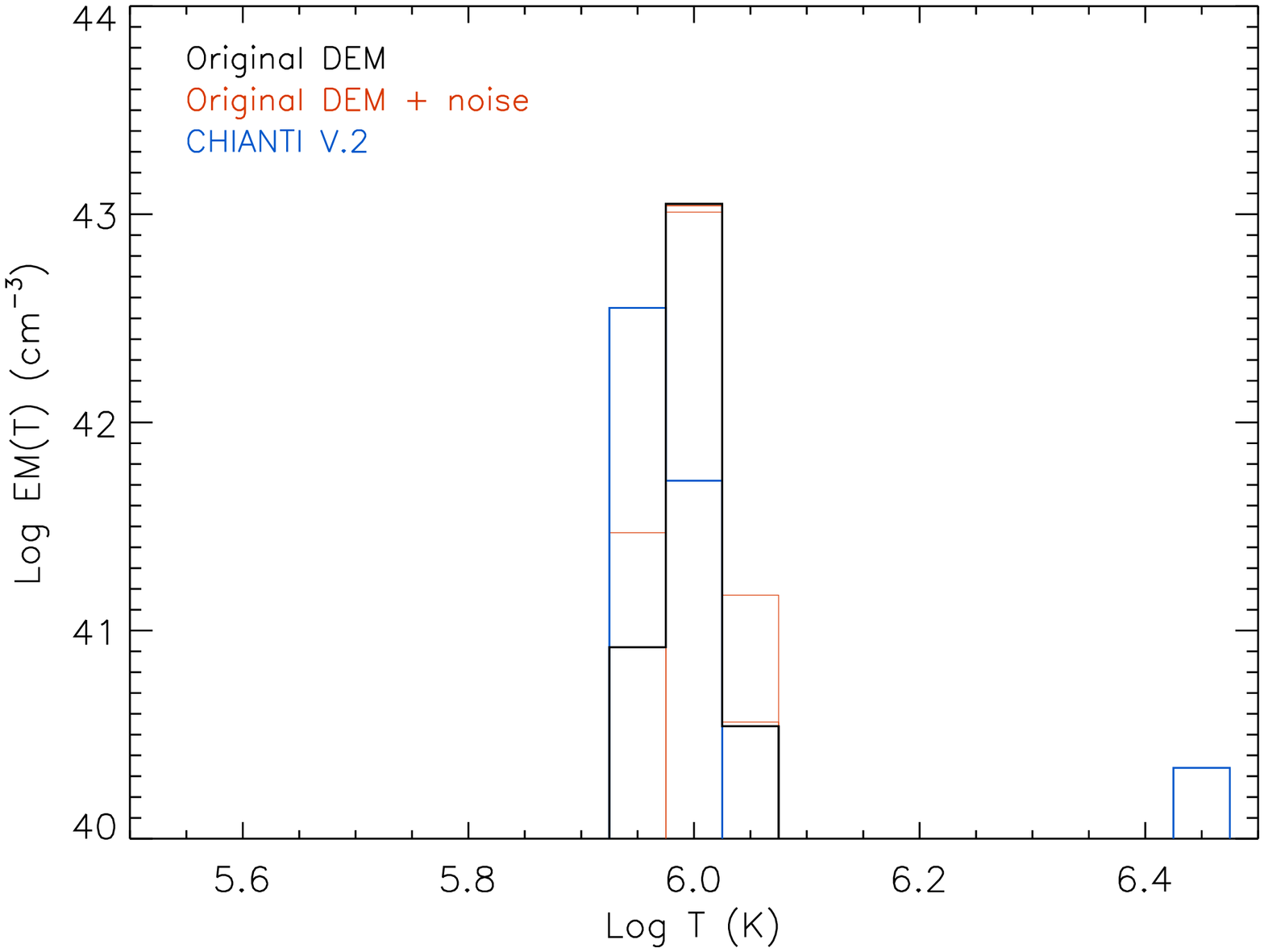}
\includegraphics[width=5.0cm,height=8.0cm,angle=90]{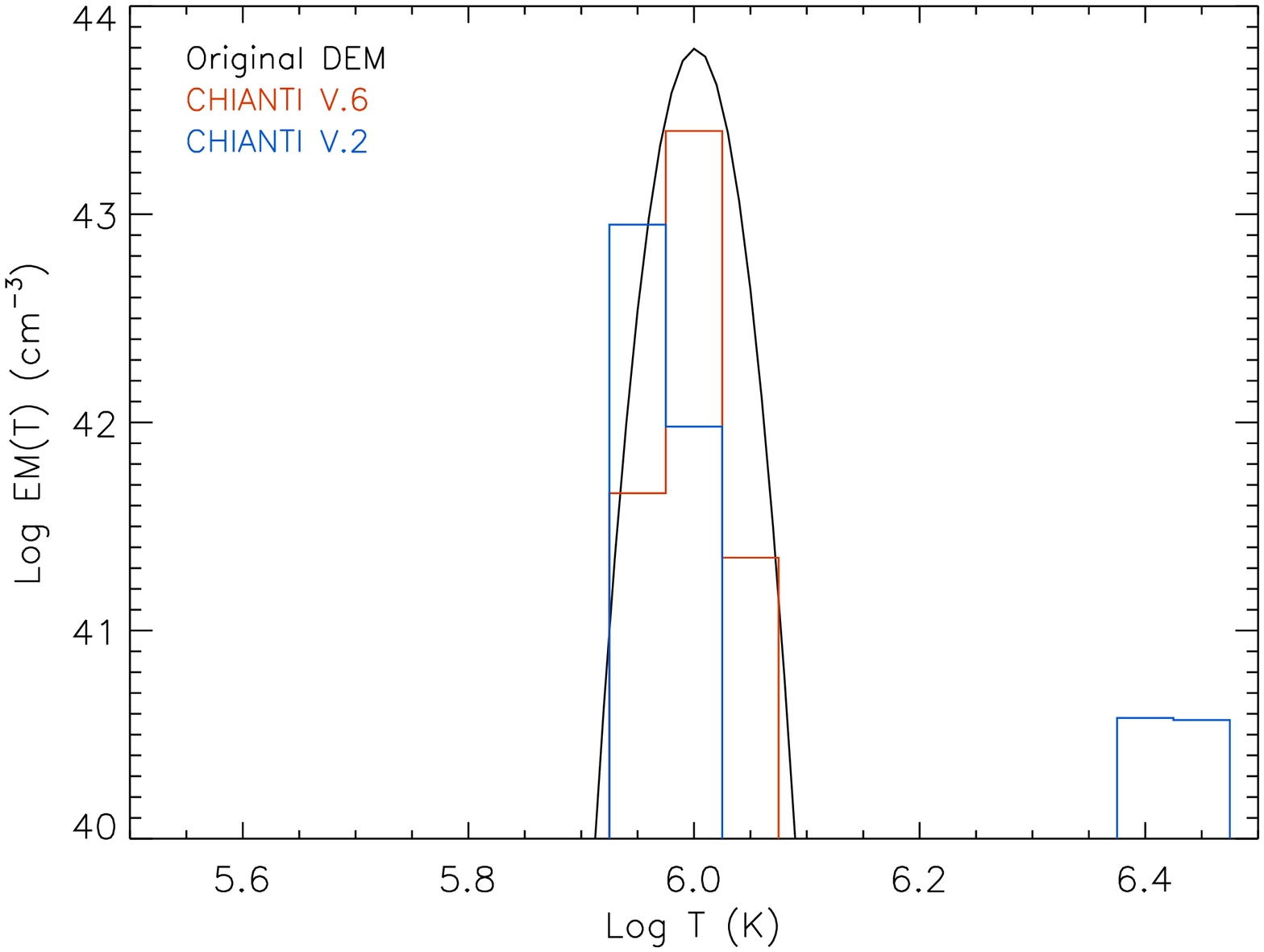}
\caption{\label{v2-a} Comparison of $EM(T)$ reconstructions of a plasma made 
of a single isothermal component (top) and a single Gaussian component (bottom); each
reconstructed using two different versions of CHIANTI: version 6.0.1 (red curves) and 
2 (blue curves). Thin lines in the top panel indicate the case with random noise
added, from Section~\ref{isot-comp}.}
\end{figure}

\begin{figure}
\includegraphics[width=5.0cm,height=8.0cm,angle=90]{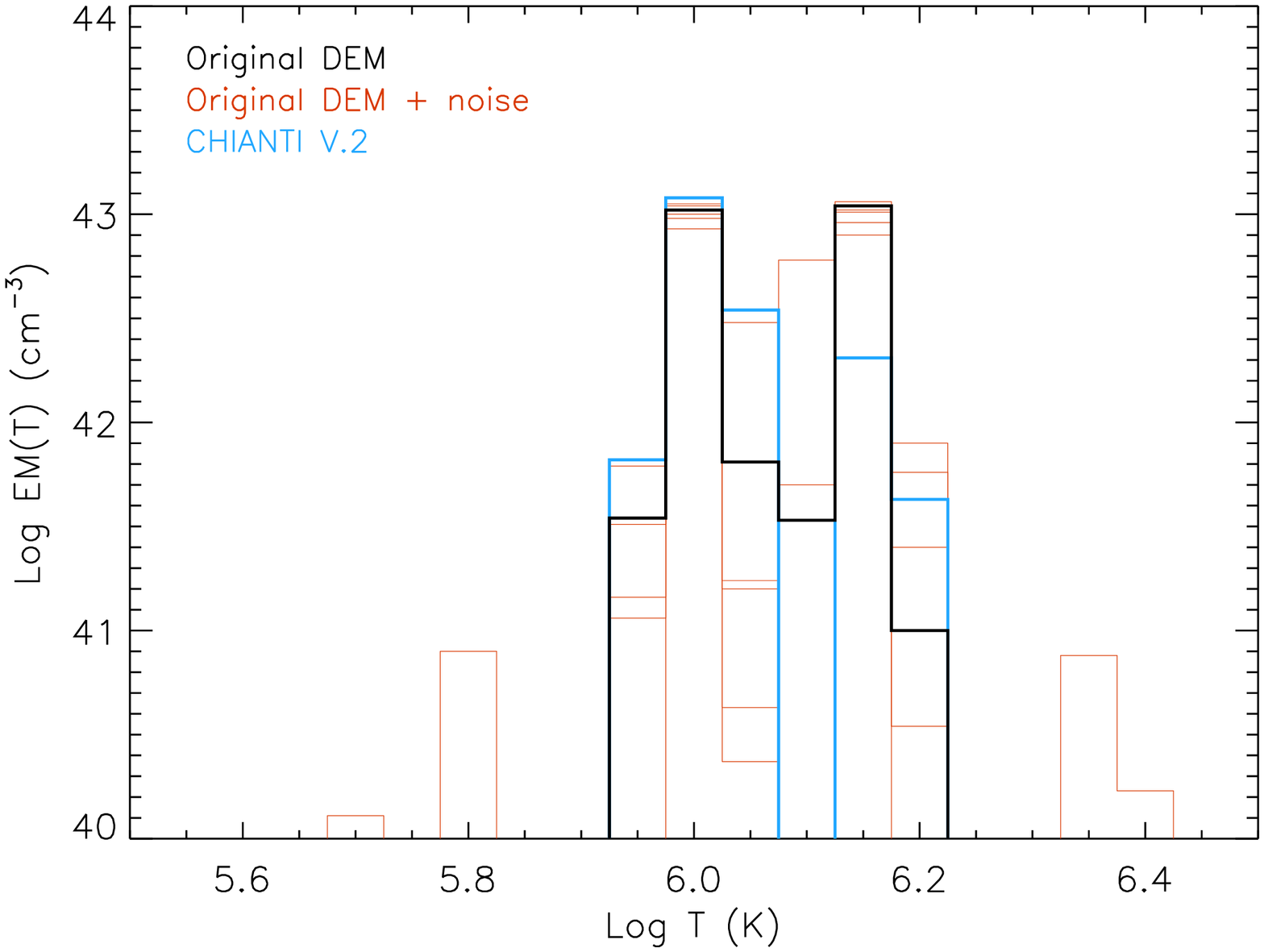}
\includegraphics[width=5.0cm,height=8.0cm,angle=90]{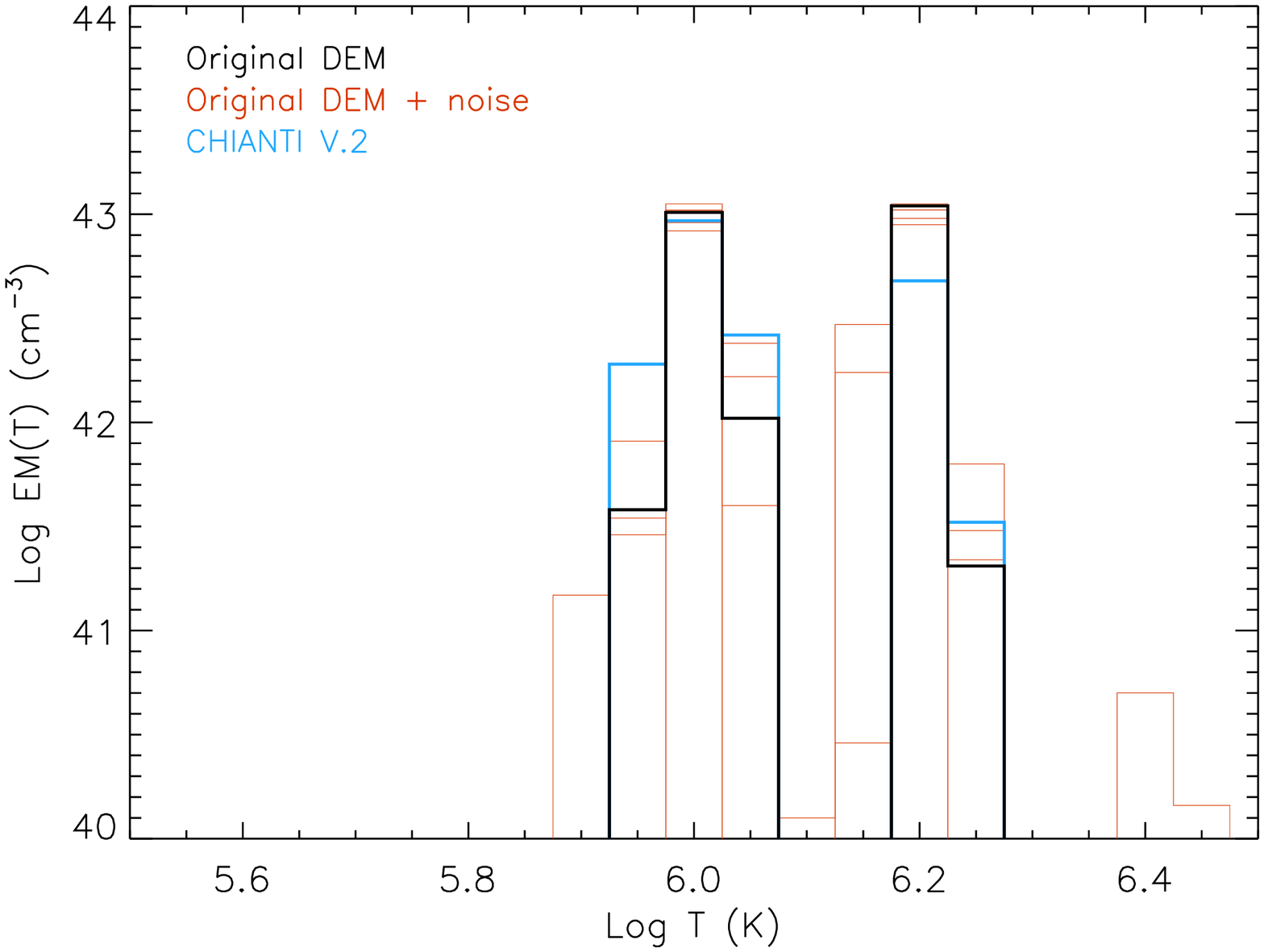}
\caption{\label{v2-b} Comparison of $EM(T)$ reconstructions of a plasma 
made of two isothermal components separated by $\Delta log T=0.15$ (top) and 
$\Delta log T=0.2$ (bottom); each reconstructed using two different versions 
of CHIANTI: version 6.0.1 (red curves) and 2 (blue curves). Thin lines in 
the top panel indicate the case with random noise added, from 
Section~\ref{isot-comp}.}
\end{figure}

Figures~\ref{v2-a} and \ref{v2-b} display (in blue) the results obtained when the emissivities 
of CHIANTI version 2 are used to determine the $EM(T)$ curves using lines calculated with
CHIANTI version 6.0.1. The red lines indicate the results obtained in the previous sections,
for comparison purposes. Figure~\ref{v2-a} illustrates the isothermal and single-Gaussian
(with $W=0.05$) cases: a single, almost identical peak is retrieved in both cases, confirming 
that the DEM method is still unable to distinguish between a true isothermal plasma and one 
with a narrow Gaussian distribution. However, one difference to be noted is that the peak
of the single component is slightly shifted towards lower temperatures. This is mostly due
to the Mazzotta \etal (1998) Fe ion abundances used with CHIANTI V.2 being shifted towards 
lower temperatures than the Bryans \etal (2009) ones used with CHIANTI V.6. The level of
noise in the V.2 solution is comparable to that of the datasets with random noise added.

Figure~\ref{v2-b} shows the case of two isothermal components, with separation $\Delta \log T=0.15$
and $0.20$. In the latter case, the two peaks are fully retrieved and separated, and the level
of noise of the solution is comparable to the noise in the solutions obtained when random
errors are included in the simulated spectra. When $\Delta \log T=0.15$, the two peaks
are less clearly separated. The temperatures of the two peaks are slightly lower than in the 
original model, but their separation is correct. The only marked difference between the solutions 
obtained with V.6 and V.2 emissivities lies in the value of the peak EM in the hot isothermal 
component, since V.2 emissivities underestimate this by a factor 5 and 2 in the $\Delta \log T=0.15$ 
and $0.20$ cases, respectively.

\section{Conclusions}
\label{conclusions}

In this work we have tested the MCMC diagnostic technique by Kashyap \& Drake (1998) 
in order to: determine its capability to reconstruct a given plasma thermal 
distribution (isothermal or multicomponent) from a set of individual line intensities; 
to study how the presence of random errors affects the MCMC reconstructions;
to investigate the optimal size of the temperature bin width $W$ as a compromise 
between retaining as much temperature structure as possible and minimizing the 
effect of noise in the solution; and to test the effect of different atomic data 
sets on the final results.
We did this by applying the MCMC technique to sets of lines whose intensities were 
calculated using a known thermal distribution first, and then randomized the intensities 
within 20\% to simulate experimental uncertainties. 

Our study shows that
the optimal bin width $W$ is $W=0.05$, as smaller values cause the MCMC technique
to provide spurious results and a high level of noise in the final reconstruction, while
larger values oversmooth the results.
Even when the emitting plasma is strictly isothermal, the MCMC technique is unable
to distinguish between a truly isothermal solution and a Gaussian DEM with FWHM=0.05;
also, the MCMC technique is able to separate multiple near-isothermal EM components only 
if their separation in temperature is $\Delta \log T=0.20$.
Atomic data uncertainties can affect the results by providing less accurate peak
EM values and shifting the EM peak temperature, but the $\Delta \log T=0.20$ resolving
power of the MCMC technique is unaffected.
The number of available ions does not affect the quality of the reconstruction, 
provided these ions are formed over a temperature range larger than the range where
the plasma EM is significant. A smaller ion formation temperature range decreases the 
temperature resolution achieved by the MCMC technique.

\bigskip

The work of Enrico Landi is supported by NASA grants NNX10AM17G and NNX11AC20G. Fabio 
Reale acknowledges support from Italian Ministero dell'Universit\'a e Ricerca and from 
Agenzia Spaziale Italiana (ASI), contract I/015/07/0. Paola Testa was supported by 
NASA contract NNM07AB07C to the Smithsonian Astrophysical Observatory. We thank the
referee for valuable comments that helped us improve the manuscript.

\end{document}